\documentclass[twocolumn,showpacs,aps,prl,superscriptaddress]{revtex4}

\usepackage{graphicx}
\usepackage{dcolumn}
\usepackage{amsmath}
\usepackage{epsfig}

\input babarsym

\newcommand{\BABARPubYear}    {08}
\newcommand{\BABARPubNumber}  {052}

\newcommand{\SLACPubNumber} {13486}

\def\figurebox#1#2#3{%
    \def\arg{#3}%
    \ifx\arg\empty
    {\hfill\vbox{\hsize#2\hrule\hbox to #2{\vrule\hfill\vbox to #1{\hsize#2\vfill}\vrule}\hrule}\hfill}%
    \else
    {\hfill\epsfbox{#3}\hfill}%
    \fi}

\begin{document}

\preprint{\babar-PUB-\BABARPubYear/\BABARPubNumber} 
\preprint{SLAC-PUB-\SLACPubNumber} 

\begin{flushleft}
\babar-PUB-\BABARPubYear/\BABARPubNumber\\
SLAC-PUB-\SLACPubNumber\\
\end{flushleft}

\title{
{\large \bf
   Evidence for the {\boldmath $\eta_b(1S)$} Meson in Radiative {\boldmath $\Upsilon(2S)$} Decay
 }
}

%% author list as of 03-Nov-2008 (481 authors)
%
\author{B.~Aubert}
\author{M.~Bona}
\author{Y.~Karyotakis}
\author{J.~P.~Lees}
\author{V.~Poireau}
\author{E.~Prencipe}
\author{X.~Prudent}
\author{V.~Tisserand}
\affiliation{Laboratoire de Physique des Particules, IN2P3/CNRS et Universit\'e de Savoie, F-74941 Annecy-Le-Vieux, France }
\author{J.~Garra~Tico}
\author{E.~Grauges}
\affiliation{Universitat de Barcelona, Facultat de Fisica, Departament ECM, E-08028 Barcelona, Spain }
% ...
\collaboration{The \babar\ Collaboration}
\noaffiliation

\date{\today}

\begin{abstract}
We have performed a search for the $\eta_b(1S)$ meson
in the radiative decay of the $\Upsilon(2S)$ resonance using a sample of $91.6 \times 10^6$ $\Upsilon(2S)$ events
recorded with the \babar\ detector at the PEP-II $B$ factory at the SLAC
National Accelerator Laboratory.
We observe a peak in the photon energy spectrum at $E_\gamma = 609.3 ^{+4.6}_{-4.5} {\rm (stat)}\pm 1.9{\rm (syst)}$ MeV, corresponding to an $\eta_b(1S)$ mass of $9394.2 ^{+4.8}_{-4.9} {\rm (stat)} \pm 2.0{\rm (syst)}$ MeV/$c^2$.
The branching fraction for the decay $\Upsilon(2S)\rightarrow\gamma\eta_b(1S)$ is determined to be
$[3.9 \pm 1.1 {\rm (stat)}  ^{+1.1}_{-0.9} {\rm (syst)}] \times 10^{-4}$. 
We find the ratio of branching fractions ${\cal B}[\Upsilon(2S)\rightarrow\gamma\eta_b(1S)]/{\cal B}[\Upsilon(3S)\rightarrow\gamma\eta_b(1S)] = 0.82 \pm 0.24 {\rm (stat)}$$ ^{+0.20}_{-0.19}{\rm (syst)}$.

\end{abstract}


\maketitle
A candidate for the $\eta_b(1S)$ meson, the ground state of the bottomonium system,
was recently observed in the radiative decays of the \ThreeS~\cite{ref:etabdiscovery}. The \babar\ experiment has accumulated a large sample of data at the peak of the \TwoS resonance, where radiative \TwoS decays are also expected to produce the $\eta_b(1S)$ meson. Theoretical predictions for ${\cal B}[\Upsilon(2S)\rightarrow\gamma\eta_b(1S)]$ range from $(1-15)\times 10^{-4}$~\cite{ref:GodfreyRosner}. 
A 90\% confidence level upper limit of ${\cal B}[\Upsilon(2S)\to \gamma\, \eta_b(1S)] <5.1\times10^{-4}$ is provided by the CLEO III experiment~\cite{ref:cleo}.

The ratio of branching fractions for the transitions $\Upsilon(2S)\rightarrow\gamma\eta_b(1S)$ and
$\Upsilon(3S)\rightarrow\gamma\eta_b(1S)$ is dependent upon the overlap integrals of the relevant bottomonium wave functions~\cite{ref:GodfreyRosner}, enabling a test that the observed state is the $\eta_b(1S)$ meson. More generally, the measured hyperfine mass splitting between the triplet and singlet states
in the bottomonium system provides a better understanding of nonrelativistic bound states in QCD and the role of spin-spin
interactions in quarkonium models~\cite{QWG-YR, ref:Gray}.

In this Letter, we report evidence for the radiative transition
$\Upsilon(2S) \to \gamma \, \eta_b(1S)$. Hereafter $\eta_b(1S)$ will be abbreviated as $\eta_b$. 

The data used in this analysis were collected with the \babar\ detector~\cite{ref:babar} 
at the PEP-II asymmetric-energy $e^+e^-$ storage rings.  
The primary data sample consists of 14 \invfb of integrated luminosity collected at the peak of the $\Upsilon(2S)$ resonance.
An additional sample of 44 \invfb collected 40~MeV below the $\Upsilon(4S)$ resonance is used for background and efficiency studies. 
The trajectories of charged particles are reconstructed
with a combination of five layers of double-sided silicon strip 
detectors and a 40-layer drift chamber, both operated in the 1.5-T magnetic field
of a superconducting solenoid.
Photons are detected with a CsI(Tl) electromagnetic calorimeter 
(EMC).
The photon energy resolution varies from 3.4\% (at 300 MeV) to 2.8\% (at 800 MeV).
Hereafter we quote values of $E_{\gamma}$ measured in the center-of-mass (c.m.) frame.

 The signal for $\Upsilon(2S) \to \gamma \, \eta_b$ is extracted 
 from a fit to the inclusive photon energy spectrum.
 The monochromatic photon from this decay should appear as a peak in the photon energy spectrum
 near 615\mev
 on top of a smooth nonpeaking background from
 continuum ($e^+e^- \to q\bar q$ with $q=u,d,s,c$) events and
 bottomonium decays.

 Two other processes produce peaks in the photon energy spectrum
 close to the signal region: ISR production of the \OneS and double radiative decays of the \TwoS.
 The second transition in the processes
 $\Upsilon(2S) \to \gamma \chi_{bJ}(1P), \chi_{bJ}(1P)\rightarrow\gamma\Upsilon(1S), \ J=0, 1, 2$, 
 produces peaks centered at 391, 423, and 442\mev, respectively.
 These three peaks are merged due to
 photon energy resolution and the small Doppler broadening that arises from the motion of the $\chi_{bJ}(1P)$ 
in the c.m.~frame.
 We use the $\chi_{bJ}(1P)\rightarrow\gamma\Upsilon(1S)$ signal to validate estimates of signal efficiencies and determine the absolute photon energy scale.
 Radiative production of the $\Upsilon(1S)$ via initial state radiation (ISR), $e^+e^- \to \gamma_{ISR} \, \Upsilon(1S)$, leads to a peak near 550\mev. 
The signal peak is better separated from the peaking background, with respect to the $\Upsilon(3S)\rightarrow\gamma\eta_b$ analysis~\cite{ref:etabdiscovery}, primarily due to better absolute energy resolution at lower energy.

Decays of the $\eta_{b}$ via two gluons, expected to be its dominant decay mode, have 
high charged-particle multiplicity. 
We select hadronic events by requiring four or more charged tracks in the event and 
that the ratio of the second to zeroth Fox-Wolfram moments~\cite{ref:fox} be less than 0.98.

Photon candidates are required to be isolated from all charged tracks.
To ensure that their EMC shapes are consistent with an electromagnetic shower,
the lateral moment~\cite{ref:LAT} is required to be less than 0.55.
To ensure high reconstruction efficiency and good energy resolution, the signal photon candidate is required to lie in the central 
angular region of the EMC, $-0.762<\cos(\theta_{\gamma, \rm{LAB}})<0.890$, 
where $\theta_{\gamma, \rm{LAB}}$ is the angle between the photon and the beam axis in the 
laboratory frame.

The correlation of the direction of the photon with the 
thrust axis~\cite{ref:brandt} of the $\eta_b$ is small, 
as there is no preferred direction in the decay of the spin-zero $\eta_b$ and
the momentum of the $\eta_b$ is small in the c.m.~frame.
In contrast, there is a strong correlation between the photon direction
and thrust axis in continuum events.
The thrust axis is computed with all charged tracks
and neutral calorimeter clusters in the event, with the exception of the
signal photon candidate.
A requirement of $|\cos{\theta_T}|< 0.8$ is made to reduce continuum background,
where $\theta_T$ is the angle between the thrust axis and the momentum of the signal photon
candidate.

A principle source of background is photons from \piz decays.
A signal photon candidate is rejected if in combination with another photon
in the event it forms a \piz candidate whose mass is within 15 MeV/$c^2$ of the nominal \piz mass.
To maintain high signal efficiency,
we require the second photon of the \piz candidate
to have an energy in the laboratory frame greater than $40~\mev$.
 
	The final efficiency evaluated from simulated events is 35.8\%.

The selection criteria were chosen by maximizing
the ratio $N_{S}/\sqrt{N_{T}}$, where $N_S$ is the signal yield and $N_T$ is the total yield of events in the signal region. 
The result of the optimization is insensitive to the exact definition of the signal region.
A detailed Monte Carlo (MC) simulation~\cite{ref:geant} provides the signal sample for this optimization, while a small fraction (7\%) of the $\Upsilon(2S)$ data is used to model the background.
To avoid a potential bias, these data are excluded from the final fit
of the photon energy spectrum. 
The remaining $\Upsilon(2S)$ data used for the analysis have an integrated luminosity of 13 \invfb,
corresponding to $(91.6 \pm 0.9)$ million $\Upsilon(2S)$ events.

To extract the $\eta_b$ signal, a $\chi^2$ fit of the $E_\gamma$ spectrum 
is performed in the region $0.27<E_\gamma<0.80$\gev.
The fit includes four components: 
nonpeaking background,
$\chi_{bJ}(1P)\rightarrow \gamma \, \Upsilon(1S)$, $\gamma_{ISR} \Upsilon(1S)$, and the $\eta_b$ signal.  

The nonpeaking background
is parametrized by an empirical probability density function (PDF)
$A\exp{\left(\sum^{4}_{i=1} c_{i} x^{i}\right)}$, 
where $x=E_{\gamma}$, and $A,c_{i}$ are determined in the fit. 

Doppler-broadened Crystal Ball (CB) functions~\cite{ref:CB} are used as phenomenological PDFs for the three 
$\chi_{bJ}(1P)\rightarrow \gamma \Upsilon(1S)$ shapes.
The CB function is a Gaussian modified to have an extended power-law tail
on the low (left) side.
 The power law parameter describing the low-side tail of the CB function
is common to all three of the $\chi_{bJ}(1P)$ peaks. The Doppler broadening of the $\chi_{bJ}(1P)$ peaks is modeled by analytically convolving the CB functions with rectangular functions of half-width 6.5, 5.5, and 4.9 \mev for the $J=0,1,2$ states, respectively. These values are evaluated using the $\Upsilon(2S)$ and $\chi_{bJ}(1P)$ masses~\cite{ref:PDG}. The resolution parameter of the $\chi_{b0}(1P)$ PDF is fixed to that of the $\chi_{b2}(1P)$. Due to its small yield and its position on the low side of the $\chi_{bJ}(1P)$ peak, the exact width of the $\chi_{b0}(1P)$ is not crucial.
The relative rates of the $\chi_{bJ}(1P)$ components are fixed to values determined 
from a control sample of $\Upsilon(2S)\rightarrow\gamma\chi_{bJ}(1P), \chi_{bJ}(1P)\rightarrow\gamma\Upsilon(1S), \Upsilon(1S) \rightarrow \mu^+\mu^-$ events,
and the relative peak positions from the world-averaged (PDG) values~\cite{ref:PDG}, with a photon energy scale offset determined in the fit.

The PDF of the peaking background from ISR $\Upsilon(1S)$ production 
is parametrized as a CB function with parameters determined from simulated events. The ISR peak position is fixed to the value determined by the \OneS and \TwoS masses~\cite{ref:PDG}, minus the energy scale offset shared with the $\chi_{bJ}(1P)$ peaks. 

\begin{figure}
\begin{center}
\includegraphics[scale=0.45]{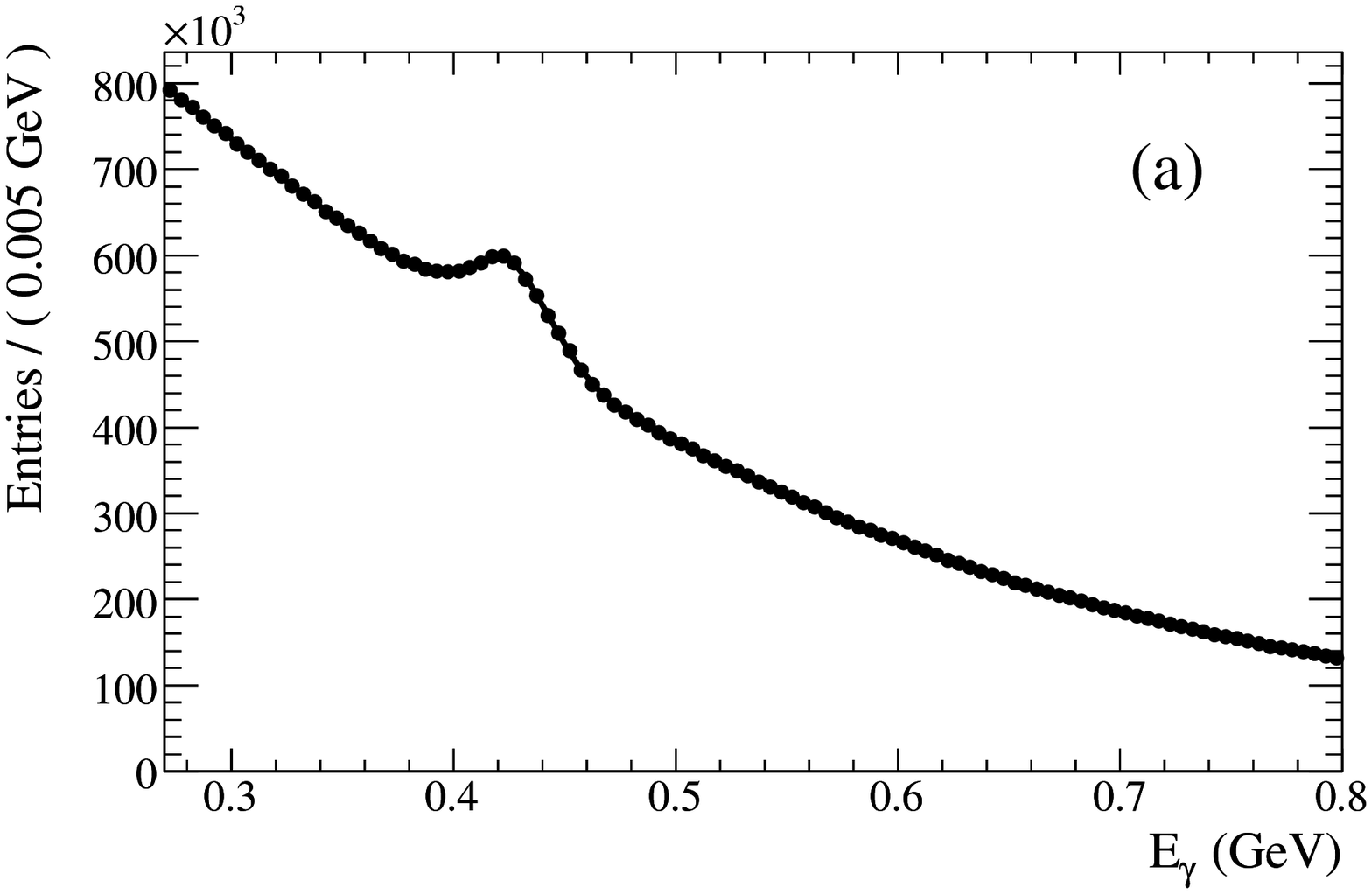}
\includegraphics[scale=0.45]{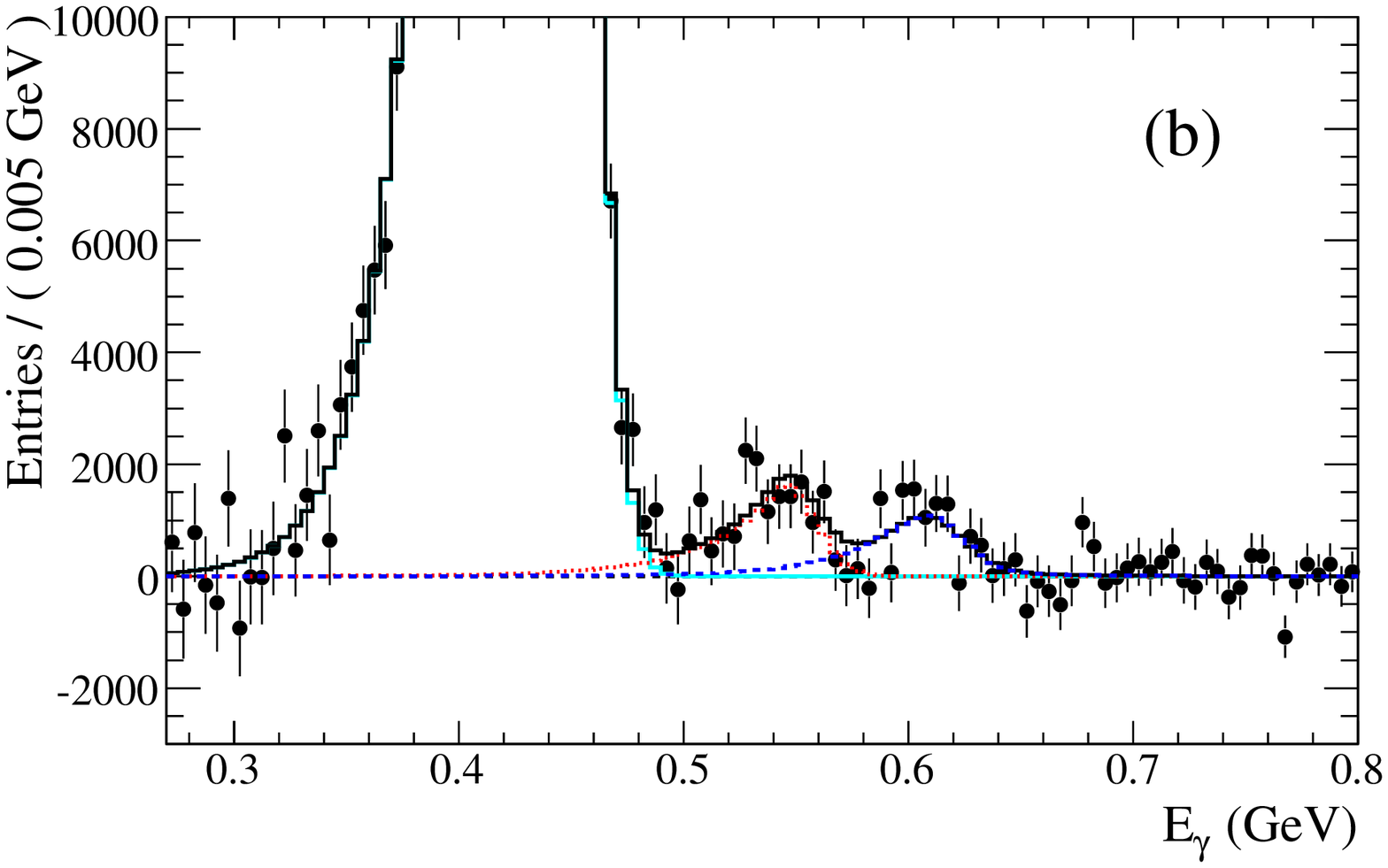}
\end{center}
\caption{\label{fig:backgroundsubtracted} 
(a) Inclusive photon spectrum in the region $0.27<E_\gamma<0.80$\gev.
    The fit is overlaid on the data points.
(b) (color online) Inclusive photon spectrum after subtracting the nonpeaking background,
with the PDFs for $\chi_{bJ}(1P)$ peak (light solid line), ISR $\Upsilon(1S)$ (dotted line), $\eta_b$ signal (dashed line) and 
the sum of all three (dark solid line). 
}
\end{figure}

The $\eta_b$ signal PDF is a nonrelativistic Breit-Wigner function convolved with
a CB function to account for the experimental $E_\gamma$ resolution.
The CB parameters are determined from signal MC.
Theoretical predictions for the $\eta_b$ width, based on the expected ratio of 
the two-photon and two-gluon widths, range from 4 to 20 MeV~\cite{ref:widththeory}.
Since the width of the $\eta_b$ is not known, we have chosen a nominal value of 10 MeV, as in the \ThreeS analysis.

The free parameters in the fit are the normalizations of all fit components, all of the nonpeaking background PDF parameters, the $\eta_b$ peak 
position, the energy scale offset, the $\chi_{b1}(1P)$ and $\chi_{b2}(1P)$ CB resolutions, and the transition point between the Gaussian and power law components of the $\chi_{bJ}(1P)$ CB functions.

Figure 1 shows the photon energy spectrum and the fit result before (a) and after (b) subtraction of the nonpeaking background.
The $\chi^2$ per degree of freedom from the fit is 115/93. 
The line shapes of the three peaking components, $\chi_{bJ}(1P)$, ISR $\Upsilon(1S)$, and the $\eta_b$ signal, are clearly visible in the subtracted spectrum. 
The $\eta_b$ signal yield is $12800 \pm 3500$ events, and the $\eta_b$ peak energy is $607.9 ^{+4.6}_{-4.5}$ MeV. 
The observed signal width is consistent with being dominated by the resolution of 18~\mev.

The ISR $\Upsilon(1S)$ yield can be estimated from data collected below the $\Upsilon(4S)$ resonance. After correcting for the luminosity ratio, and the difference in ISR cross section and detection efficiencies at the two energies, we expect $16700 \pm 700 \pm 1200$ ISR $\Upsilon(1S)$ events in the on-resonance \TwoS data sample. 
The consistency of the observed yield of the $\Upsilon(1S)$ component, $16800^{+4200}_{-4000}$ events, with the expected value provides an important validation of the fitted background rate near the signal region. 
The yield and peak position of the $\eta_b$ signal change by less than $0.1\sigma$ 
when the ISR $\Upsilon(1S)$ yield is fixed to the expected value.
 
We estimate the systematic uncertainty by varying
the Breit-Wigner width  in the $\eta_b$ PDF to 5, 15, and 20 MeV,
varying the PDF parameters fixed in the fit by $\pm$1 $\sigma$, using alternative smooth background shapes, varying the histogram binning between 1 and 15~MeV,  incorporating a high-side tail to the $\chi_{bJ}(1P)$ peaks, and subtracting possible peaking background components. 
Smooth background PDF variations consist of using alternative smooth background shapes that either incorporate a 3rd-order polynomial in the exponential (i.e. $c_{4}=0$) or use a PDF of the form $k\left(E_{\gamma}/E_0\right)^{-\Gamma_1} [1 + (E_{\gamma}/E_0)^{1/\alpha}]^{-\left(\Gamma_2-\Gamma_1\right)\alpha}$.
Other background shape variations consisting of adding a term $c_5 x^5$ to the exponential of the smooth background function or adding a constant background PDF were found to change the fit negligibly. 
An additional high-side tail in the $\chi_{bJ}(1P)$ peak may be produced by the coincidental overlap of photons from $\chi_{bJ}(1P)$ decays with particles from the rest of the event or beam debris. 
We model this tail as a 90\mev wide Gaussian centered about each of the $\chi_{bJ}(1P)$ peaks. 
Due to the large width of this component, it is indistinguishable from the nonpeaking background, and its inclusion does not improve the fit. We take the difference between the nominal fit and the fit including this tail as a systematic error. 
To evaluate the systematic due to the $\chi_{b0}(1P)$ resolution, we perform a fit in which the $\chi_{b0}(1P)$ resolution is fixed to that of the $J=1$ state.
To investigate the possible effect of peaking background from $\Upsilon(2S)\rightarrow(\pi^0,\eta)\Upsilon(1S)$ events, these contributions are subtracted prior to fitting, assuming the measured value and 90\% CL upper limits for the branching fraction of the $\eta$ and $\pi^0$~\cite{ref:PDG} transitions, respectively, giving a variation of $-71$ ($+651$) events for the $\eta$ ($\pi^0$) transition.
Photons from the transition $\Upsilon(2S)\rightarrow\pi^0\pi^0\Upsilon(1S)$, which produce a smoothly varying background below 400~MeV, are absorbed into the smooth background PDF and do not require a separate systematic error.
  
Including systematic uncertainties, the signal yield is $12800 \pm 3500$$^{+3500}_{-3100}$ events. 
The largest contributions to the systematic error on the $\eta_b$ yield are from the $\eta_b$ width variation ($^{+1700}_{-1200}$ events) and the background shape variation ($^{+2600}_{-2700}$ events). 

The photon energy scale is corrected with the fitted energy scale offset of $ 1.4 \pm 0.2 \pm 0.7 $ MeV determined from the $\chi_{bJ}(1P)$ and ISR peaks.
The systematic error is half of the shift added in quadrature with the PDG errors on the $\chi_{bJ}(1P)$ masses (0.4\mev).
The ISR peak contributes negligibly to the determination of the offset, due to its small yield.
After including an additional systematic uncertainty of 1.7 MeV from the fit variations described above, we obtain a value of
$E_\gamma = 609.3  ^{+4.6}_{-4.5} \pm 1.9$ MeV for the $\eta_b$ signal. 

To confirm that this state is identical to the state observed in the $\Upsilon(3S)\rightarrow\gamma\eta_{b}$ analysis~\cite{ref:etabdiscovery} we calculate the significance of the signal with the signal peak fixed to 614.3\mev, the value expected for an $\eta_b$ mass of 9388.9\mev. The $\eta_b$ signal significance is estimated as $\sqrt{\chi^2({\rm no\ signal})-\chi^2({\rm fixed\ mass})}$, where $\chi^2({\rm fixed\ mass})$ is the $\chi^2$ of the fit with the $\eta_b$ signal included and $\chi^2({\rm no\ signal})$ is the $\chi^2$ of the fit with the $\eta_b$ PDF removed. The statistical significance estimated in this way is 3.7 standard deviations.
The significance of the signal, including systematics, is estimated by making the variations discussed above. 
Additional cross-checks are performed by changing the lower (upper) limit of the fit range to 250~MeV (850~MeV) and varying the selection on $|\cos(\theta_T)|$.
In all fits, the significance lies between 3.0 and 4.3 standard deviations.

The $\eta_b$ mass derived from the $E_\gamma$ signal is
$M(\eta_b) = 9394.2 ^{+4.8}_{-4.9} \pm 2.0$ MeV/$c^2$.
Using the PDG value of $9460.3 \pm 0.3$ MeV/$c^2$ for the $\Upsilon(1S)$ mass,
we determine the $\Upsilon(1S)$-$\eta_b$ mass splitting to be $66.1 ^{+4.9}_{-4.8} \pm 2.0$ MeV/$c^2$.

For the measurement of the branching fraction, we have an additional source of uncertainty resulting from the signal selection efficiency.  
The systematic uncertainty on the photon detection efficiency is 1.8\%.
 We estimate the uncertainty on the hadronic selection efficiency (4.9\%) by comparing data and MC efficiencies of the selection on hadronic $\Upsilon(1S)$ events. 
The uncertainty in photon quality selection efficiency (0.5\%) is estimated from \piz decays in data and MC.
The difference between the efficiency in MC and the efficiency for a flat distribution (0.6\%) is used as the uncertainty on the $|\cos{\theta_T}|$ selection. 
We determine the uncertainty for the $\pi^0$ selection (4.1\%) by comparing the efficiency-corrected $\chi_{bJ}(1P)$ yield with and without the \piz veto.
The total systematic error on the selection efficiency is 6.7\%. The uncertainty on the number of \TwoS events is 0.9\%. 
Incorporating these systematic uncertainties, we determine the branching fraction of the decay 
$\Upsilon(2S) \to \gamma \, \eta_b$ to be $(3.9 \pm 1.1 ^{+1.1}_{-0.9} ) \times 10^{-4}$.

In the \ThreeS analysis~\cite{ref:etabdiscovery}, we estimated the systematic uncertainty on the signal efficiency using $\chi_{bJ}(2P)$ decays, incurring a large error (22\%) due to the uncertainties in the $\chi_{bJ}(2P)$ branching fractions. 
The uncertainty in $\Upsilon(3S)\rightarrow\gamma\eta_b$ efficiency obtained using the procedure described above is 5.5\%,  resulting in a final branching fraction of $\mathcal{B}[\Upsilon(3S)\rightarrow\gamma\eta_b]=(4.8\pm0.5\pm0.6)\times 10^{-4}$. 
This value supersedes our previous result, which differs only in having a systematic uncertainty two times larger.

Using the results given above, we determine a branching fraction ratio of ${\cal B}[\Upsilon(2S)\rightarrow\gamma\eta_b]/{\cal B}[\Upsilon(3S)\rightarrow\gamma\eta_b] = 0.82 \pm 0.24$$ ^{+0.20}_{-0.19}$. 
The systematic uncertainties due to selection efficiency and the unknown $\eta_b$ width partially cancel in the ratio. 
Our measurement is consistent with some of the theoretical estimates of this ratio of magnetic dipole transitions
to the $\eta_b$, $0.3 - 0.7$~\cite{ref:GodfreyRosner}, 
while the absolute transition rates are not well-predicted by theoretical models.

In conclusion, we have obtained evidence, with a significance of 3.0 standard
deviations, for the radiative decay of the $\Upsilon(2S)$ to a narrow state with a mass
slightly less than that of the $\Upsilon(1S)$. The ratio of the radiative production rates for this state at the \TwoS and \ThreeS resonances is consistent with that expected of the $\eta_b$. Under this interpretation, the mass of the $\eta_b$ is $9394.2 ^{+4.8}_{-4.9} \pm 2.0$ MeV/$c^2$, which corresponds to a mass splitting between the $\Upsilon(1S)$ and the $\eta_b$ of 
$66.1^{+4.9}_{-4.8} \pm 2.0$ MeV/$c^2$, consistent with the value from the $\Upsilon(3S)$ analysis. 
The average of the two results is $M(\eta_b) = 9390.8 \pm 3.2$ MeV/$c^2$. 
This value of the $\eta_b$ mass is consistent with a recent unquenched lattice prediction~\cite{ref:Gray} but more than two standard deviations away from the mass predicted by approaches based on perturbative QCD~\cite{ref:pQCD}.

We are grateful for the excellent luminosity and machine conditions
provided by our \pep2\ colleagues, 
and for the substantial dedicated effort from
the computing organizations that support \babar. 
The collaborating institutions wish to thank 
SLAC for its support and kind hospitality. 
This work is supported by
DOE
and NSF (USA),
NSERC (Canada),
CEA and
CNRS-IN2P3
(France),
BMBF and DFG
(Germany),
INFN (Italy),
FOM (The Netherlands),
NFR (Norway),
MES (Russia),
MEC (Spain), and
STFC (United Kingdom). 
Individuals have received support from the
Marie Curie EIF (European Union) and
the A.~P.~Sloan Foundation.


\begin{thebibliography}{99}

\bibitem{ref:etabdiscovery} B.~Aubert {\it et al.} (\babar\ Collaboration), Phys. Rev. Lett. {\bf 101}, 071801 (2008); {\bf 102}, 029901(E) (2009).

\bibitem{ref:GodfreyRosner}
S.~Godfrey and J.~L.~Rosner, Phys. Rev. D {\bf 64}, 074011 (2001); {\bf 65}, 039901(E) (2002), and references therein.

\bibitem{ref:cleo}
M.~Artuso {\it et al.} (CLEO III Collaboration), Phys. Rev. Lett. {\bf 94}, 032001 (2005).

\bibitem{QWG-YR} For a comprehensive review, see N.~Brambilla {\it et al.}  
(Quarkonium Working Group), CERN Yellow Report, CERN-2005-005 (2005).

\bibitem{ref:Gray}
   A.~Gray {\it et al.} (HPQCD and UKQCD Collaborations), Phys.\ Rev.\  D {\bf 72}, 094507 (2005); 
   T.~Burch and C.~Ehmann, Nucl. Phys. {\bf A797}, 33(2007);
   T.-W.~Chiu {\it et al.} (TWQCD Collaboration), Phys. Lett. B {\bf 651}, 171 (2007).

\bibitem{ref:babar}
B.~Aubert {\em et al.} (\babar\ Collaboration), \nima{479}, 1 (2002).

\bibitem{ref:fox} G.~C.~Fox and S.~Wolfram, Nucl. Phys. {\bf B149}, 413 (1979).

\bibitem{ref:LAT} A.~Drescher {\em et al.} (ARGUS Collaboration), \nima{237}, 464 (1985). 

\bibitem{ref:brandt} S.~Brandt {\em et al.}, Phys. Lett. {\bf 12}, 57 (1964);
E.~Farhi, Phys. Rev. Lett. {\bf 39}, 1587 (1977).

\bibitem{ref:geant} The \babar\ detector Monte Carlo simulation is based on GEANT4: S.~Agostinelli {\it et al.}, \nima{506}, 250 (2003); 
Simulated events are generated with Jetset 7.4: T.~Sj$\rm \ddot{o}$strand and M. Bengtsson, Comput. Phys. Commun. {\bf 43}, 367
(1987).

\bibitem{ref:CB}
J.~E.~Gaiser, Ph.D. thesis, Stanford University [SLAC-R-255] (1982). 

\bibitem{ref:PDG}
C.~Amsler {\em et al.} (Particle Data Group), \plb{667}, 1 (2008).

\bibitem{ref:widththeory} 

   W.~Kwong, P.~B.~Mackenzie, R.~Rosenfeld, and J.~L.~Rosner, Phys. Rev. D {\bf 37}, 3210 (1988);
   C.~S.~Kim, T.~Lee, and G.~L.~Wang, \plb{606}, 323 (2005);
   J.~P.~Lansberg and T.~N.~Pham, Phys. Rev. D {\bf 75}, 017501 (2007).

\bibitem{ref:pQCD} 
B.~A.~Kniehl {\em et al.}, Phys. Rev. Lett. {\bf 92}, 242001 (2004);
S.~Recksiegel and Y.~Sumino, \plb{578}, 369 (2004).

\end{thebibliography}
\end{document}